\documentclass[twocolumn,11pt]{aastex63}
\usepackage{amsmath,amstext}
\usepackage[T1]{fontenc}
\usepackage{apjfonts} 
\usepackage[figure,figure*]{hypcap}
\received{~--- xxx}
\revised{~--- xxx}
\accepted{~--- xxx}

\shorttitle{Compact star-forming galaxies: Bluebery}
\shortauthors{Paswan et al.}

\begin{document}


\title{SDSS-IV MaNGA: Blueberry Candidates Associated with LSB Galaxies $-$ Merger or Tidal Dwarf Systems ?}




\correspondingauthor{Abhishek Paswan}
\email{paswanabhishek@iucaa.in}

\author{Abhishek Paswan}
\affil{Inter-University Centre for Astronomy and Astrophysics, Ganeshkhind, Post Bag 4, Pune 411007, India}

\author{Kanak Saha}
\affiliation{Inter-University Centre for Astronomy and Astrophysics, Ganeshkhind, Post Bag 4, Pune 411007, India}

\author{Suraj Dhiwar}
\affiliation{Dayanand Science College, Latur 413512, India}


\begin{abstract}

We report here our finding of two new blueberry galaxies using optical IFU spectroscopic data from the MaNGA survey. Both the blueberries are found to be compact ($\textless$ $1 - 2$ kpc) starburst systems located at the outskirts of Low Surface Brightness (LSB) galaxies. Our blueberries have the lowest stellar masses $\sim$ 10$^{5}$ M$_{\odot}$ amongst the locally known blueberry galaxies.
We find a significantly large mean metallicity difference ($\sim$ 0.5 dex) between the blueberry sources and their associated LSBs. Moreover, the radial metallicity gradients in our blueberries are also different than their respective LSBs - suggesting that these had significantly different metallicity histories. The likelihood of survival of these blueberries as TDGs is analyzed based on their structural and kinematic properties. Our analysis shows that although the two blueberries are stable against internal motions, they would not have survived against the tidal force of the host galaxy. Based on the velocity difference between the host LSBs and the blueberries, it appears that they are compact, starburst systems in their advanced stage of merger with these LSBs situated in a dense environment. Implications of our findings are discussed.


   
\end{abstract}


\section{Introduction} \label{sec:intro}

One of the most important events in the history of our early Universe is the cosmic re-ionization, which is still observationally not well-constrained due to limited sample of observed ionizing sources. In recent years, a lot of efforts have been therefore made to search for the sources responsible for the re-ionization of the Universe at high redshifts z $\gtrsim$ 6. The general outcome of these studies suggests that faint, low-mass, compact and clumpy starburst systems with an escape of ionizing radiation above $10 - 20$\% are the primary sources to the ionizing budget of the early Universe \citep[e.g.,][]{Ouchi2009,Robertson2013,Dressler2015,Finkelstein2019}.\\

However, direct observation of such objects at high redshifts are difficult due to various reasons e.g., due to limitation of observational facilities available at the present day, faintness of the sources \citep{Robertson2013,Bouwens2015,Alavi2020}, contamination by low redshift interlopers \citep[e.g.,][]{Vanzella2010,Vanzella2012,Siana2015} and attenuation of ionizing photons by neutral intergalactic medium \citep[IGM;][]{Inoue2014}. As a consequence of these limitations, so far a very few reliable high-redshift compact starburst galaxies with leaking ionizing photons are known e.g., $Ion2$, $ion3$, Q1549-C25 and A2218-Flanking, AUDFs-01 as studied in \citet{Vanzella2015,Vanzella2018,deBarros2016,Shapley2016,Bian2017} and \citet{Saha2020}. But there has been a significant increase in the number of sources, in particular the Green Peas \citep[GPs;][]{Cardamone2009} and blueberries \citep{Yang2017}, with leaking ionizing radiation with an escape fraction of $2 - 98$\% in the low-redshift ($z \leq 0.4$) Universe \citep[e.g.,][]{Izotov2016a,Izotov2016b,Izotov2018a,Izotov2018b,Jaskot2019,Wang2019}. These sources are low-mass, compact ($\textless$ 1 kpc), starburst systems commonly seen with extreme blue optical color, high star formation rate (SFR; $0.05 - 100$ M$_{\odot}$~yr$^{-1}$), high EW[OIII] ($500 - 1500$ \AA), high ionization parameter (O$_{32}$ $\equiv$ [OIII]/[OII] $\geq$ 5), low dust extinction (E($B - V$) = $0 - 0.25$) and low gas-phase metallicity (12 + log(O/H) = $7.6 - 8.25$). All these characteristics are now being broadly used as vital criteria to search for compact starburst galaxies exhibiting escape of ionizing radiations \citep{Stasinska2015,Izotov2016a,Izotov2016b,Yang2017,Izotov2018a,Izotov2018b}. Nevertheless, they are still statistically less in number limiting our understanding of their physical nature in general. Several questions such as their formation scenario and spatially resolved properties remain unanswered.\\

Previous studies of GPs and blueberries have explicitly shown that they reside in low galaxy-density environments \citep{Cardamone2009,Yang2017}. This might be the case due to their selection effect, as they are primarily selected based on their visual inspection as compact green and blue isolated galaxies in the Galaxy Zoo catalogue. But their environment dependence is yet to be understood. Of these two class of galaxies, blueberries are relatively nearby objects with $z < 0.05$ and provide us with a better likelihood to explore their spatially resolved properties. To that, we perform a systematic search for possible blueberry candidates in the Mapping Nearby Galaxies at Apache Point Observatory \citep[MaNGA;][]{Bundy2015} Integral Field Unit (IFU) survey. In this paper, we report on two new blueberry galaxies with lowest stellar mass known to date and they are physically located at the outer most region of their associated low-surface-brightness (LSB) galaxies. In the rest of the paper, we discuss their possible formation mechanism and spatially resolved physical properties using MaNGA IFU data.\\

The paper is organized as follows: Section~\ref{sec:data} describes the MaNGA observation and our source selection criteria. In section~\ref{sec:host}, we discuss the host galaxy properties and the selection of H{\sc ii} regions in it. Section~\ref{sec:physical_property} describes the photometric and spectroscopic properties of the blueberries. In section~\ref{sec:metallicity}, we show the metallicity and its gradient while in section~\ref{sec:stability}, we study their stability properties. Section~\ref{sec:formation} deals with the formation scenario. The summary and conclusions are presented in section~\ref{sec:summary}. Throughout this work, we have considered the flat cosmological parameters of \textit{H$_{o}$} = 70 km s$^{-1}$ Mpc$^{-1}$, $\Omega_{m}$ = 0.3 and $\Omega_{\Lambda}$ = 0.7.

\begin{figure*}
\begin{center}
\rotatebox{0}{\includegraphics[width=0.85\textwidth]{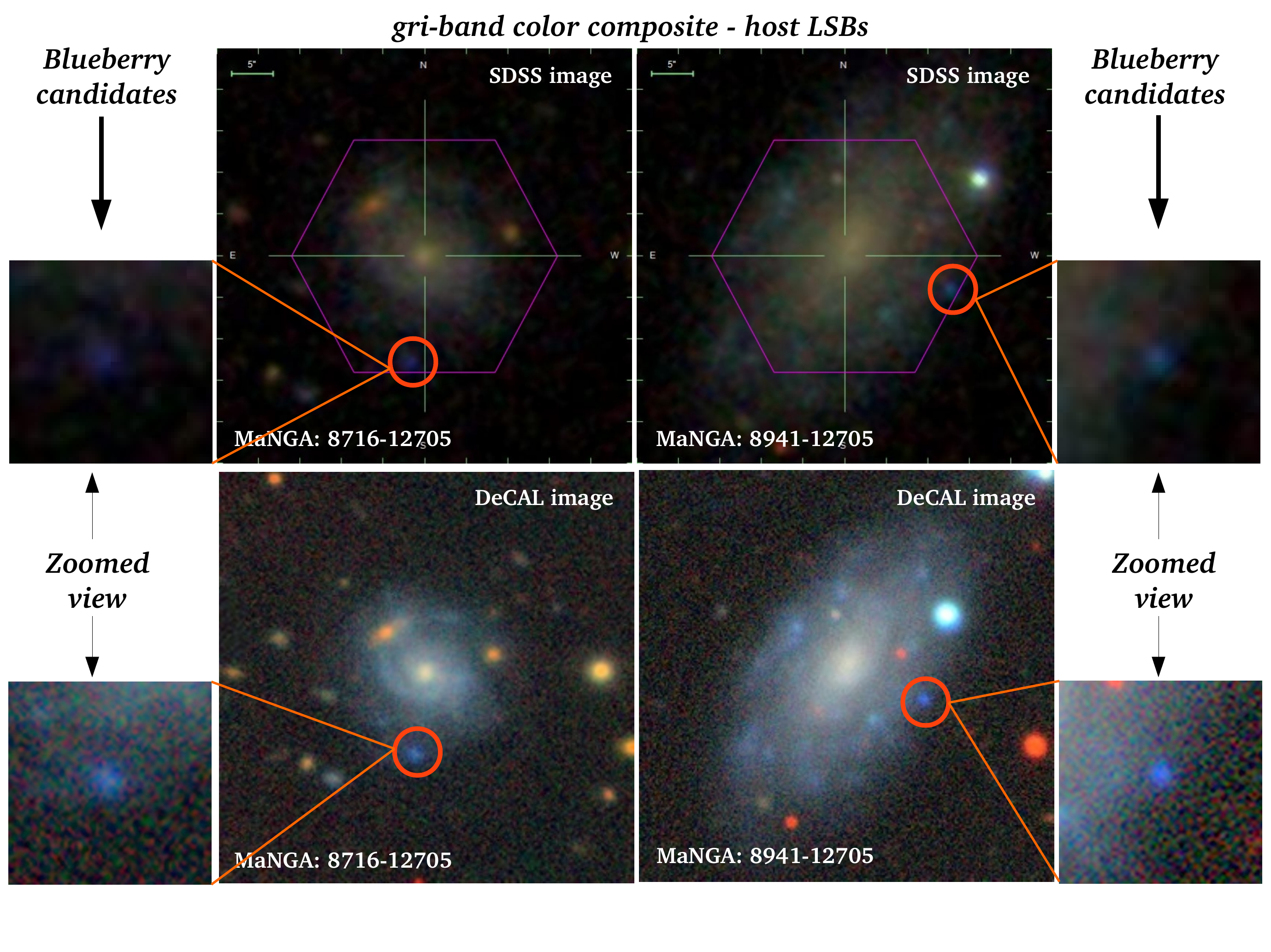}}
\caption{Upper panel shows the MaNGA footprint overlaid on the SDSS-$gri$ color composite images of galaxies in the present study. Lower panel shows the corresponding higher-resolution deep color composite images in the DeCAL survey. The identified blueberry sources are marked by red circles whose zoomed views in each panel are also shown.}
\label{color-comp}
\end{center}
\end{figure*} 

\begin{table}
\centering
\caption{Summary of the galaxy properties studied in this work.}
\vspace {0.3cm}
\begin{tabular}{ccc} \hline 
MaNGA ID  &  $8716 - 12705$  & $8941 - 12705$	  \\
Other name &  SDSS J081236.03+524403.0      & UGC 4236  \\ \toprule
RA              &  08$^{h}$ 12$^{m}$ 36.03$^{s}$ &  08$^{h}$ 07$^{m}$ 57.80$^{s}$ \\ 
Dec             &  +52$^{d}$ 44$^{m}$ 03$^{s}$   &  +26$^{d}$ 01$^{m}$ 43$^{s}$ \\
Redshift (z)    & 0.0403                         & 0.0141    \\
m$_{g}$ (mag)   & 21.808                         & 21.901 \\
m$_{r}$ (mag)   & 22.933                         & 23.126 \\
m$_{i}$ (mag)   & 22.599                         & 24.264 \\
$r_{equ}$ (kpc) & 2.14                           &   0.74 \\
$r_{eff}$ (kpc) & 1.00                           &   0.37 \\
$R$\tablenotemark{a} (kpc)  &   10.39            &  3.67 \\
$\sigma_{*}$ (km~s$^{-1}$)  & 101                & 67 \\
$M_{*}^{Blueberry}$ (M$_{\odot}$) & 2.48 $\times$ 10$^{5}$          & 1.05 $\times$ 10$^{5}$ \\
$M_{*}^{host~LSB}$ (M$_{\odot}$) & 3.2 $\times$ 10$^{9}$ & 4.4 $\times$ 10$^{9}$ \\
$M_{tidal}$ (M$_{\odot}$) & 8.39 $\times$ 10$^{7}$  & 1.08 $\times$ 10$^{8}$ \\
EW(H$\alpha$) [\AA] &   332                 &  206 \\
L(H$\alpha$)[erg~s$^{-1}$]  &   1.10 $\times$ 10$^{40}$ & 1.53 $\times$ 10$^{39}$ \\
SFR [M$_{\odot}$~yr$^{-1}$] &  0.087        & 0.012 \\
12 + log(O/H)   & 8.05                           & 8.07 \\\hline 
\end{tabular} 
\flushleft
\tablenotetext{a}{Projected distance between blueberries and their respective host LSB galaxies}
\label{Table1}
\end{table}

\section{Data and source selection}
\label{sec:data}

The data used in this work are mainly taken from 16$^{th}$ data release (DR16) of the MaNGA survey. This survey is an ongoing optical IFU observing program under the fourth generation Sloan Digital Sky Survey \citep[SDSS-IV;][]{Bundy2015}. It uses the same spectrograph as used in the Baryon Oscillation Spectroscopic Survey \citep[BOSS;][]{Smee2013} mounted on 2.5-m Sloan Foundation Telescope \citep{Gunn2006} at Apache Point Observatory (APO). Here the IFU size is selected such that it covers 1.5 R$_{e}$ of the galaxy extent \citep{Law2016}. This survey has a target to observe about 10,000 nearby (0.01 $\textless$ z $\textless$ 0.15) galaxies having stellar masses above 10$^{9}$ M$_{\odot}$ \citep{Wake2017}. The observed spectra have a wavelength coverage of $3600 - 10300$ \AA~with a spectral resolution of $R$ $\sim$ 2000 and an instrumental velocity resolution of $\sigma$ $\sim$ 60 km~s$^{-1}$. The MaNGA survey provides the data with an effective spatial resolution of  2.5" full width at half maximum (FWHM). The observed raw data are reduced and calibrated using the Data Reduction Pipeline \citep[DRP;][]{Law2016}. The science ready data are then provided after analysing the DRP's output products using Data Analysis Pipeline \citep[DAP;][]{Westfall2019}.  \\

In this study, we have used the DAP products given in the MaNGA DR16. Here DAP uses the pPXF code \citep{Cappellari2004} along with MILES stellar library to fit both the stellar continuum and spectral line features present in the spectra. In this fitting, single Gaussian function is used to model the individual emission and absorption features and provide the emission and absorption line fluxes after stellar continuum subtraction. It also gives stellar and gas kinematics parameter such as $v_{stellar}$, $\sigma_{stellar}$, $v_{gas}$ and $\sigma_{gas}$. All the emission line fluxes used in the present study are corrected for both the Galactic and internal extinctions using corresponding reddening $E(B - V)$ values. First Galactic foreground reddening correction was applied by assuming reddening law provided in \citet{O'Donnell1994}. Then internal reddening correction to the target galaxy was applied using the flux ratio of f$_{H\alpha}$/f$_{H\beta}$ emission lines by assuming the expected theoretical value of 2.86 (i.e., the Case-B recombination \citep{Osterbrock1989} with an electron temperature of $\sim$ 10$^{4}$ K and electron density of 100 cm$^{-3}$). For some spaxels, the flux ratio of f$_{H\alpha}$/f$_{H\beta}$ emission lines was found less than the expected theoretical value of 2.86. A low value of f$_{H\alpha}$/f$_{H\beta}$ is often associated with intrinsically low reddening, and hence we assumed an internal $E(B - V)$ values as zero for such cases. \\ 

The MaNGA DR16 contains a total of $\sim$ 4600 galaxies out of which two galaxies are presented here for the current study. These galaxies are selected from our search program of blueberry candidates in the MaNGA survey. Our primary spectroscopic search criteria to find the potential blueberry candidates was as follow.

\begin{itemize}
    \item We selected galaxies hosting H{\sc ii} gas emitting regions (as confirmed through their H$\alpha$ emission maps) with their peak [OIII]/[OII] emission line ratio of spaxel's as equal or above 5 (i.e., [OIII]/[OII] $\geq$ 5) in the MaNGA DR16 survey, which also simultaneously satisfy two other criteria as mentioned below.  
    
    \item The peak H$\beta$ emission line equivalent width of spaxel's is high, EW(H$\beta$) $\geq$ 100 \AA, so that the selected objects justify the presence of young starbursts with ages of $3 - 5$ Myr.
    
    \item Along with above criteria, the peak [OIII] $\lambda$5007 emission line equivalent width of spaxel's is also high, EW([OIII] $\lambda$5007) $\geq$ 500 \AA. 
    
\end{itemize}

It is to note here that the above spectroscopic selection criteria are similar to that which were previously applied by \citet{Cardamone2009} and \citet{Yang2017} for spectroscopic selection of GPs and blueberries, respectively. Although they also used photometric criteria, we have also checked the same photometric criteria for our spectroscopically selected sources to confirm them as blueberry sources (see Section.~\ref{sec:photmetry} for more details). With our search criteria as mentioned above, we found a total of 50 galaxies satisfying the first criteria. Out of these 50 galaxies, we found only three potential blueberry candidates satisfying the last two criteria. Of these three candidates, one blueberry candidate was found as an isolated dwarf system and is presented somewhere else by Paswan et al. (2020a,b; in preparation). It is important to note that this identified isolated blueberry candidate has already been characterized as LAE (Lyman-$\alpha$ Emitter) with showing a Ly$\alpha$ escape fraction of $\sim$ 10\% as obtained using direct ultra-violate (UV) observations with the Cosmic Origins Spectrograph (COS) on board $Hubble~Space~Telescope~(HST)$. Whereas, two other potential blueberry candidates are separately selected and presented here. Because they are located at the outer most region of large LSB galaxies. Therefore, they may provide the possible clue about the formation scenario of blueberry sources. \\ 

Apart from the MaNGA data, other ancillary data are taken from the publicly available SDSS survey. As per our requirements, the obtained science ready data are further analyzed using various standard packages available in Python, SExtractor and Image Reduction and Analysis Facilities (IRAF) to derive our final results which are presented below. 

\section{Host galaxy property and H{\sc ii} regions}
\label{sec:host}

\begin{figure*}
\begin{center}
\rotatebox{0}{\includegraphics[width=0.495\textwidth]{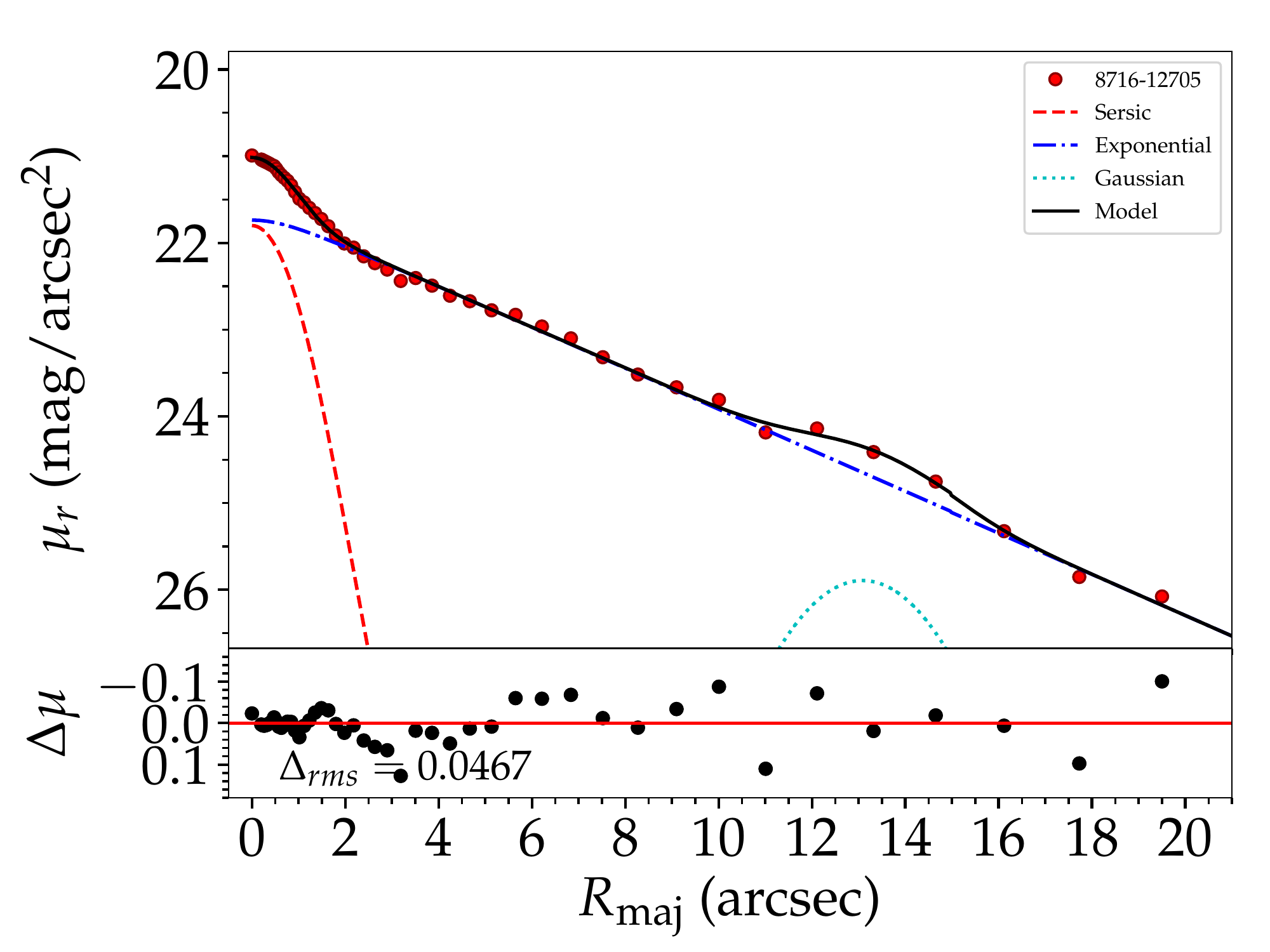}}
\rotatebox{0}{\includegraphics[width=0.495\textwidth]{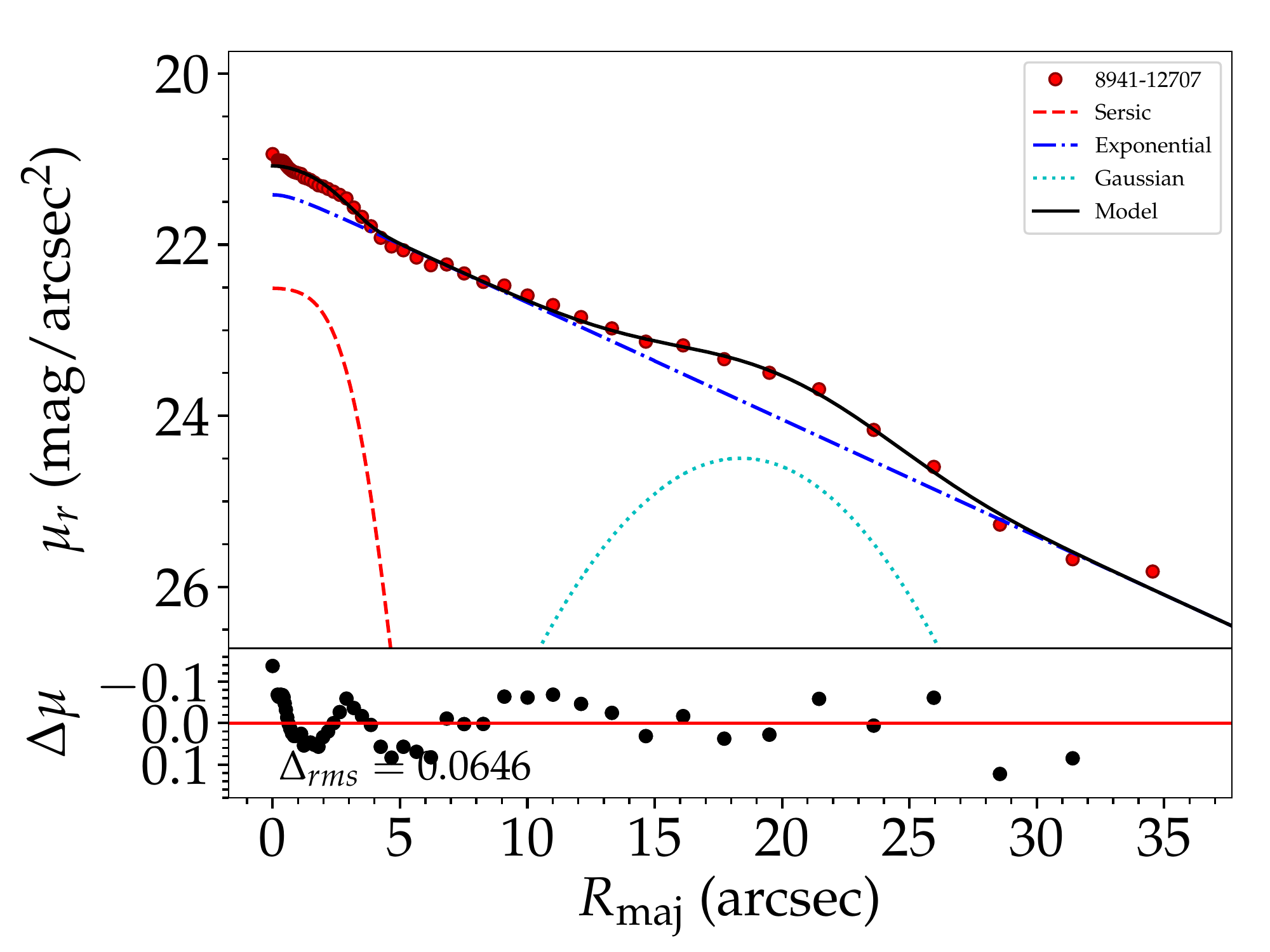}}
\caption{The Surface brightness profiles of host galaxies MaNGA: $8716 - 12705$ (left) and MaNGA: $8941 - 12705$ (right). The details of different fitted profiles are provided by legends.}
\label{SBF}
\end{center}
\end{figure*} 

In order to first characterize the nature of host galaxies as shown in Fig.~\ref{color-comp}, we have presented their $SDSS$ $r-band$ surface brightness profiles in Fig.~\ref{SBF}. In this figure, solid circles represent the observed profile derived using ELLIPSE fitting routine available in IRAF. The dashed and dot-dashed lines are the two (bulge + disk) components fitted to the observed profiles using Sersic and Exponential functions, respectively. Additionally, some of spiral features seen in the host galaxies are fitted with Gaussian models as shown by dotted-line. The solid line shows the total models fit to the observed profiles. All these fittings are performed under the PROFILER environments \citep{Ciambur2016}, after taking the image PSF (point spread function $-$ moffat) convolution into account. This provides the value of central surface brightness of disk components ($\mu_{0, disk}$) as 21.554$\pm$0.003 (MaNGA: $8716 - 12705$) and 21.310$\pm$0.005 mag arcsec$^{-2}$ (MaNGA: $8941 - 12705$). It is to note here that these values are fainter than 21 mag arcsec$^{-2}$. As the criteria used for classification of a galaxy into an high-surface-brightness (HSB) or LSB based on its $\mu_{0, disk}$ suggests that a disk galaxy can be considered as an LSB, if its $\mu_{0, disk}$ in the $r$-band is 21 mag arcsec$^{-2}$ or fainter than the quoted value \citep{Brown2001,Adami2006,Pahwa2018}. According to this criteria, the host galaxies presented here are therefore LSB galaxies.

\subsection{Selection of H{\sc ii} region of interest}

\begin{figure*}
\begin{center}
\rotatebox{0}{\includegraphics[width=0.45\textwidth]{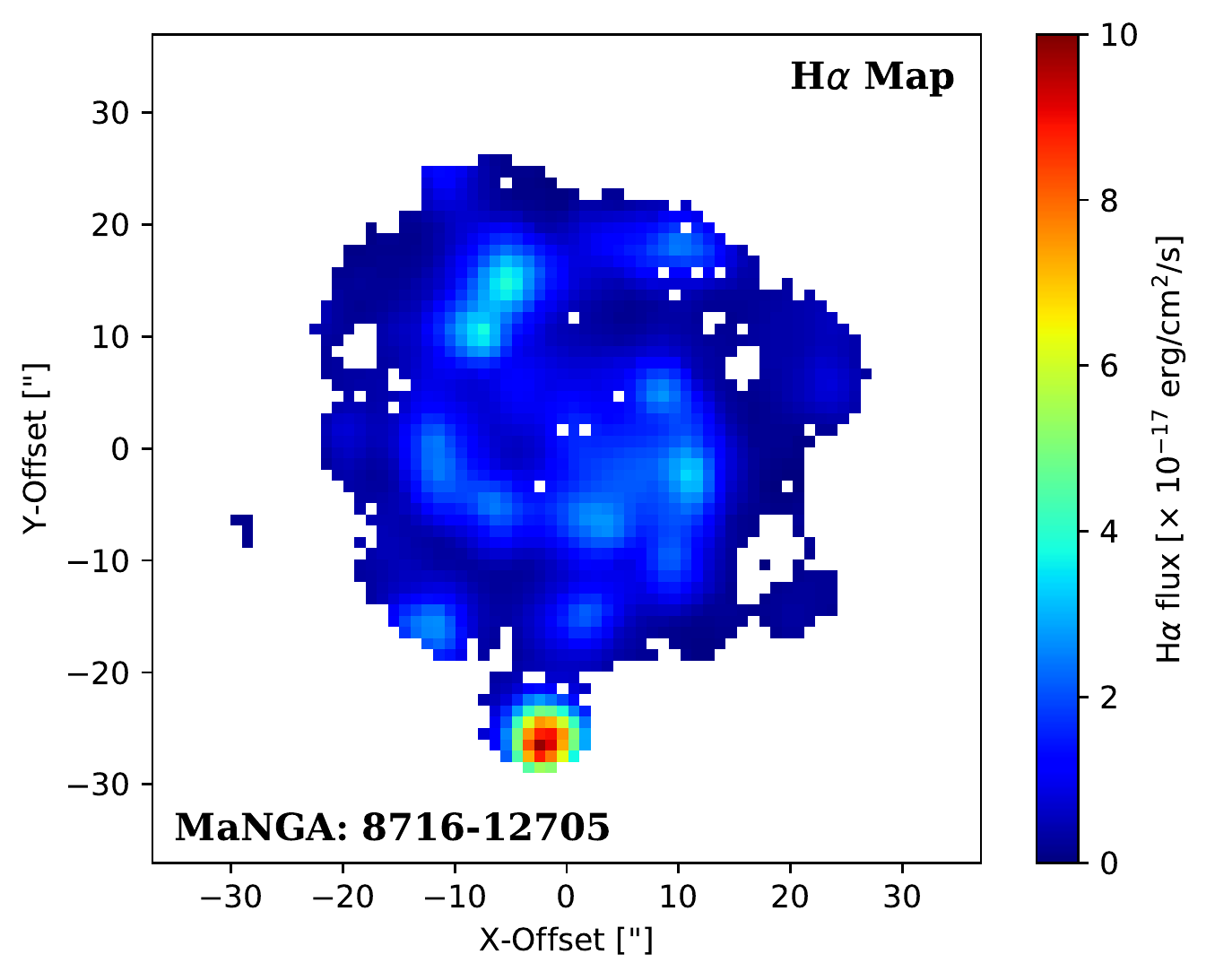}}
\rotatebox{0}{\includegraphics[width=0.45\textwidth]{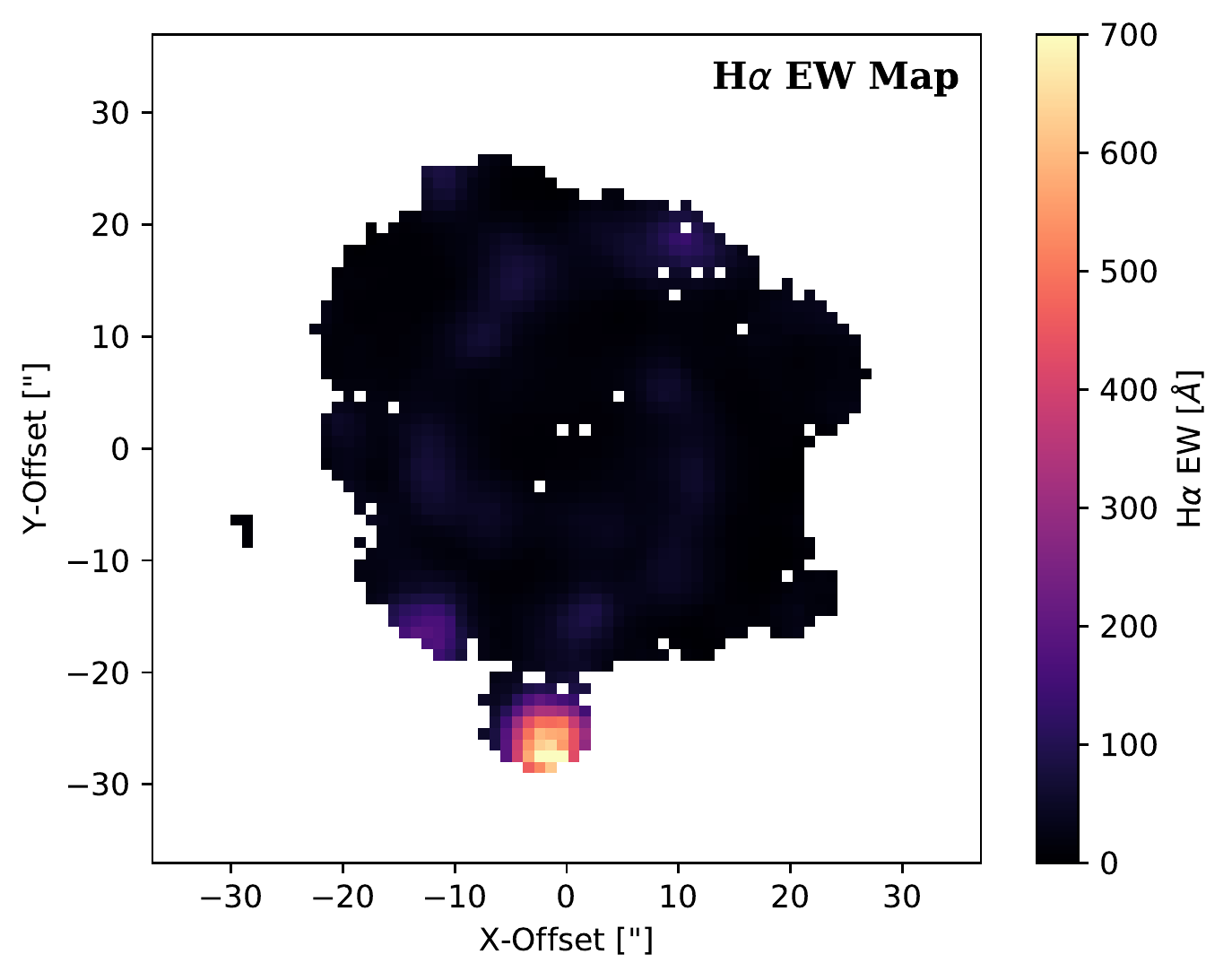}}
\rotatebox{0}{\includegraphics[width=0.45\textwidth]{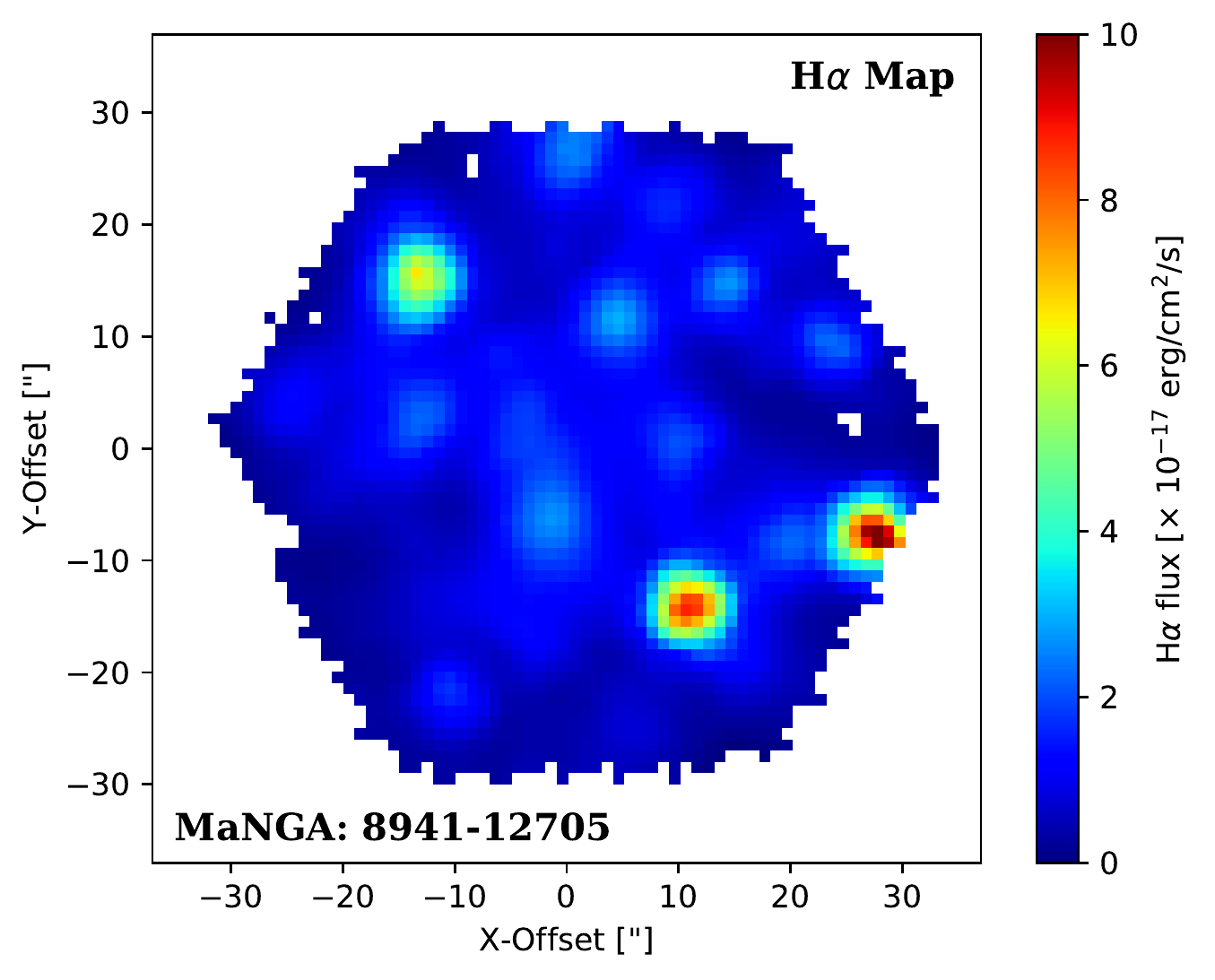}}
\rotatebox{0}{\includegraphics[width=0.45\textwidth]{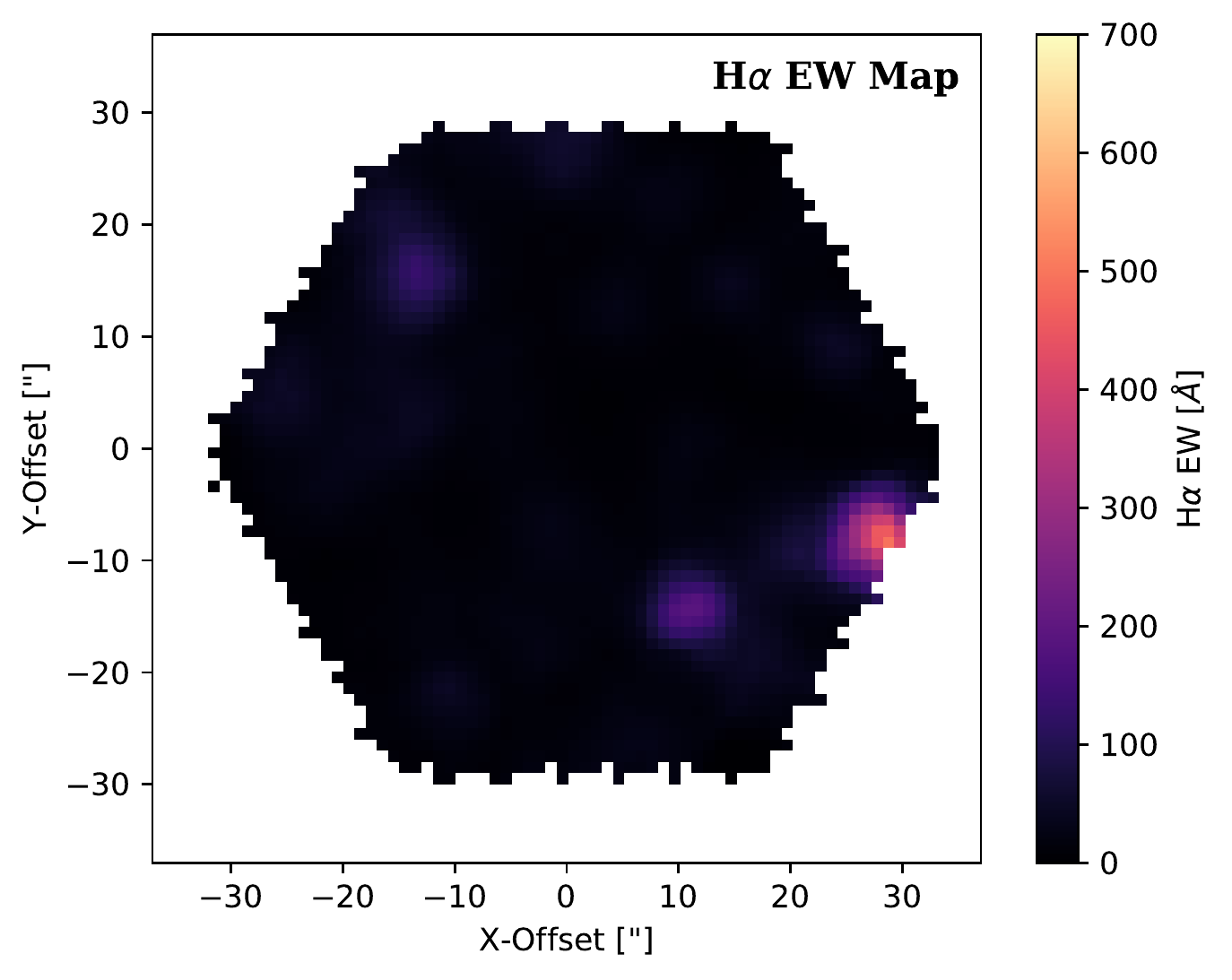}}
\caption{First column shows the H$\alpha$ emission line maps of the galaxies chosen for our study. Similarly, second column shows the equivalent width maps of H$\alpha$ emission line.}
\label{HII-sel}
\end{center}
\end{figure*}

\begin{figure*}
\begin{center}
\rotatebox{0}{\includegraphics[width=0.49\textwidth]{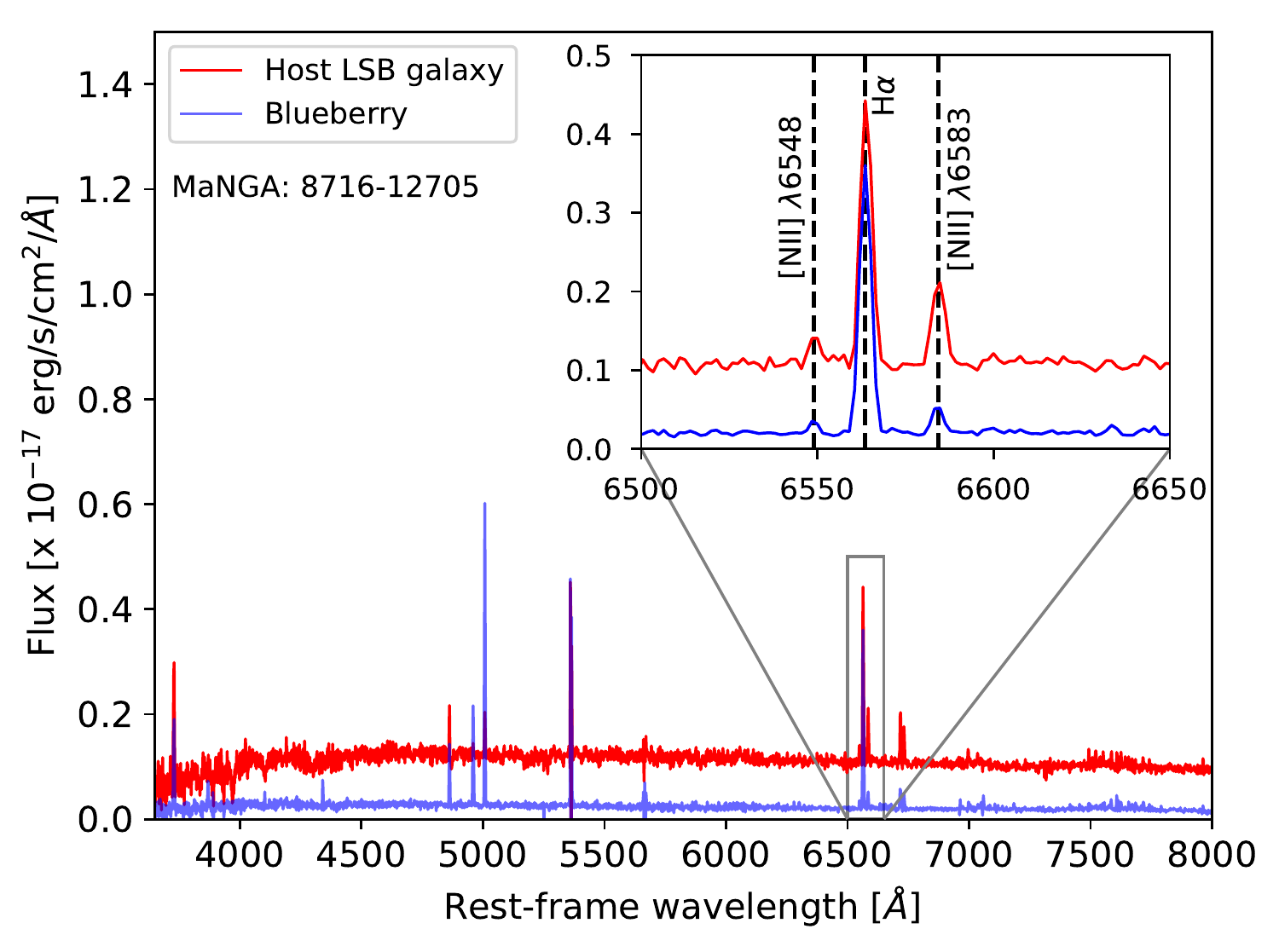}}
\rotatebox{0}{\includegraphics[width=0.49\textwidth]{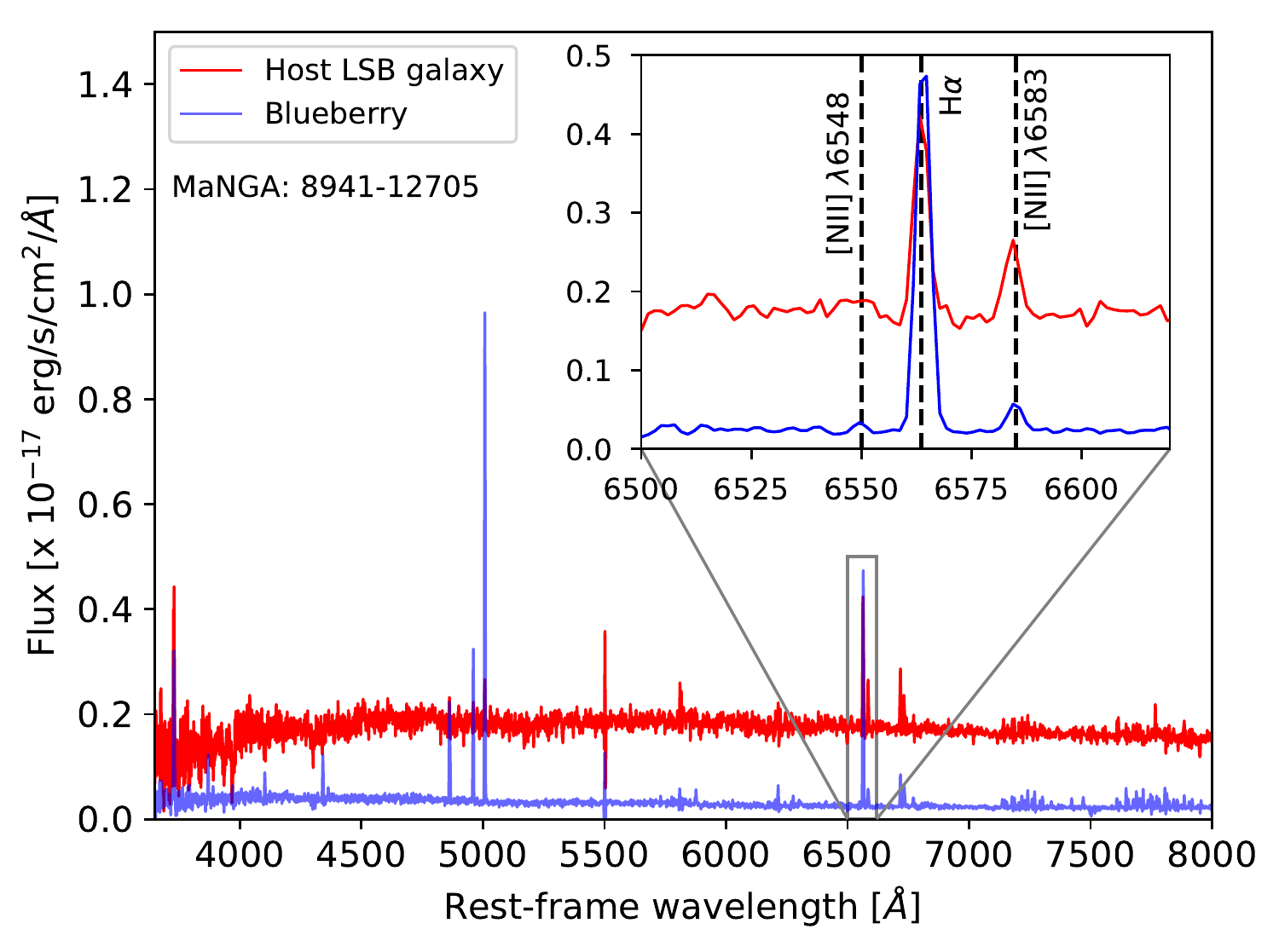}}
\caption{The central optical spectra of blueberry (blue) and host LSB galaxies (red). The inset plot in each panel represents the wavelength comparison of identified H$\alpha$+[NII] emission lines from the host LSBs and blueberries.}
\label{spec}
\end{center}
\end{figure*}

In the presented LSBs, it can be clearly seen that they host several H{\sc ii} regions which are appeared as blue regions in the color composite images (see Fig.~\ref{color-comp}), and also as bright H$\alpha$ emitting knots in the H$\alpha$ emission maps (see first column in Fig.~\ref{HII-sel}). Of all these H{\sc ii} regions, our regions of interest are selected as the brightest region in the H$\alpha$ emission line, having the largest H$\alpha$ emission line equivalent width (EW) and situated at the outer most regions of associated LSB galaxies. As such H{\sc ii} regions at outskirts of large galaxies are often identified as merging and/or tidally interacting low mass compact starburst galaxies \citep{Duc1998,Meurer1995,Melo2005,Bastian2006}. In Fig.~\ref{HII-sel}, second column shows the EW maps of H$\alpha$ emission line of the galaxies in our study. The H$\alpha$ emission line and its EW clearly indicate that the marked H{\sc ii} regions (as shown in Fig.~\ref{color-comp}) have the largest H$\alpha$ flux and EW amongst all the H{\sc ii} regions seen in LSB galaxies. These H{\sc ii} regions are also situated at the outer most region of LSB galaxies. In Fig.~\ref{spec}, we have shown the rest-frame optical spectra of these H{\sc ii} regions at outskirt (blue; marked ones in Fig.~\ref{color-comp}) and LSB galaxies at center (red), after applying the same redshift correction as measured for LSB galaxies. Both of these spectra are closely compared in the inset images of Fig.~\ref{spec}. In both the cases, the rest-frame H$\alpha$ and [NII]$\lambda\lambda$6548,6583 emission lines in the outskirt H{\sc ii} regions and LSB galaxies fall at similar wavelengths having a small radial velocity difference of $\sim$ $30$ km~s$^{-1}$, indicating that the selected outer most H{\sc ii} regions are indeed closely associated with their respective host LSB galaxies.

\section{Physical properties of the blueberries}
\label{sec:physical_property}

\subsection{Photometric characteristic: color-color diagram}
\label{sec:photmetry}

\begin{figure}[!h]
\begin{flushleft}
\rotatebox{0}{\includegraphics[width=0.48\textwidth]{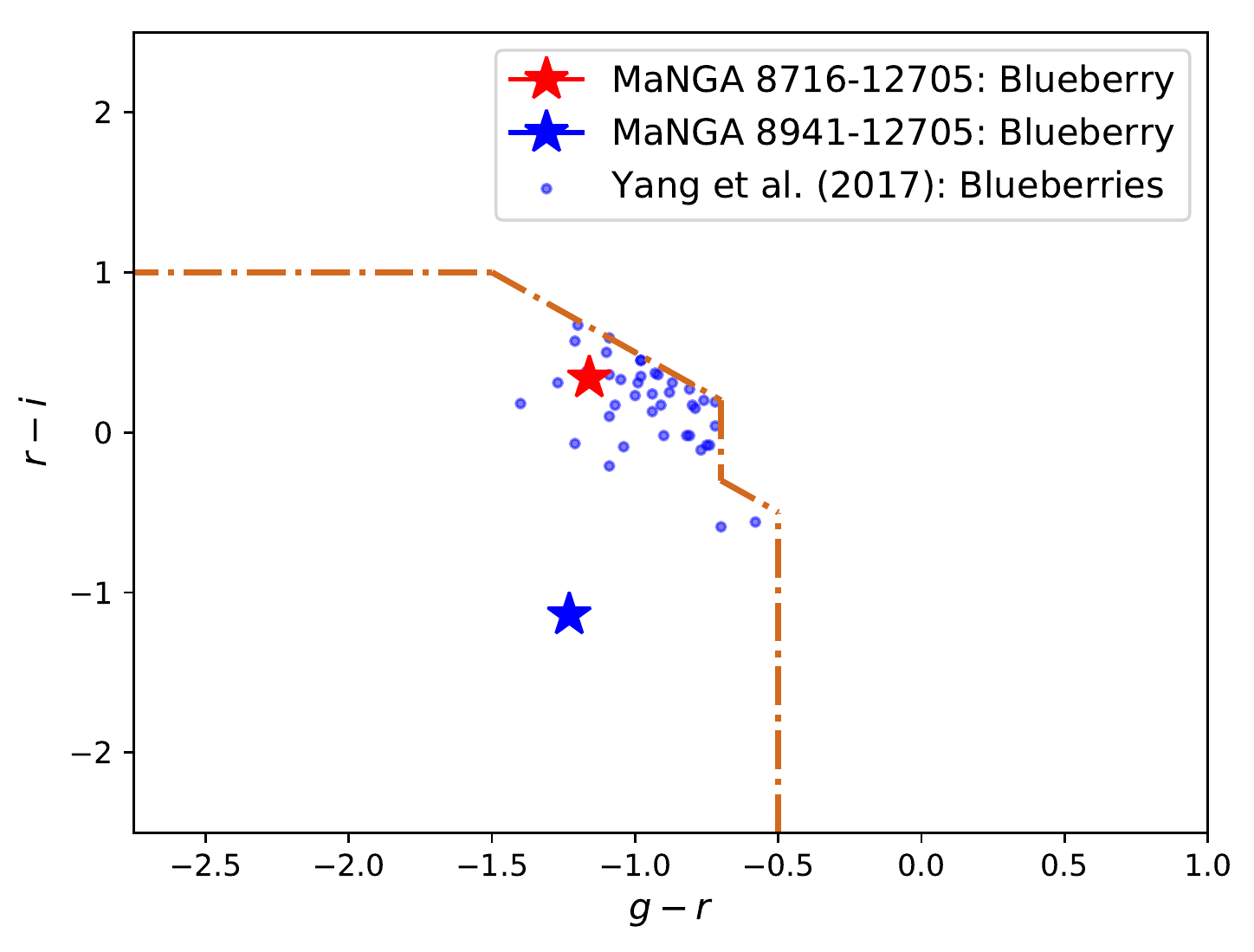}}
\caption{The $g - r$ vs $r - i$ color-color diagram of blueberry sources identified in this work as shown by star symbol. The blue dots are blueberry galaxies from \citet{Yang2017}. The orange dot-dashed line represents the color-color boundary defined for selecting blueberry galaxies in \citet{Yang2017}.}
\label{color-color}
\end{flushleft}
\end{figure}

Following the SDSS $g - r$ vs. $r - i$ color-color criteria as defined by \citet{Yang2017} to classify the blueberry candidates at lower redshift (z $\leq$ 0.05), in Fig.~\ref{color-color}, we have shown a similar color-color diagram for the selected outer most compact H{\sc ii} regions as marked by circle in color-composite images of LSB galaxies. In this figure, the previously classified blueberry galaxies and defined color boundaries in \citet{Yang2017} are also shown by blue solid circle and orange dash-dotted line, respectively. The compact H{\sc ii} regions in the present work are shown by star symbol. Interestingly, the location of the selected H{\sc ii} regions on the color-color plot are in consistent with the previous identified blueberry galaxies, showing their location inside of the color boundaries defined for classifying the blueberry galaxies. This implies that the selected compact H{\sc ii} regions are potential blueberry sources. This hypothesis is also verified by their spectroscopic properties as presented below. \\

\subsection{Spectroscopic characteristic}

\begin{figure*}
\begin{center}
\rotatebox{0}{\includegraphics[width=1.0\textwidth]{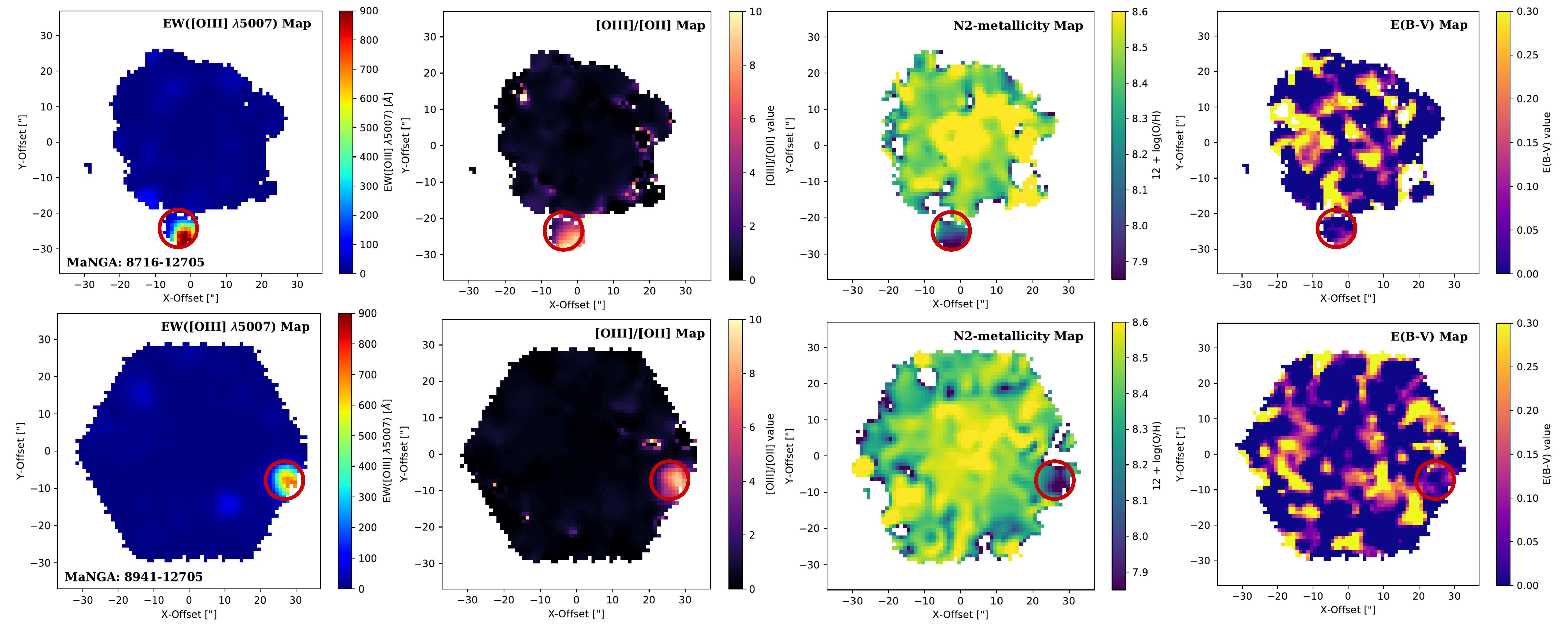}}
\caption{The spatially resolved 2D-map of EW[OIII] (first column), O$_{32}$ a proxy for ionization parameter (second column), gas-phase metallicity [12 + log(O/H)] (third column) and E($B - V$) (fourth column)  of each studied galaxy. In each panel, the identified blueberry sources are marked by red circles.}
\label{prop-map}
\end{center}
\end{figure*} 

\begin{figure*}
\begin{center}
\rotatebox{0}{\includegraphics[width=0.95\textwidth]{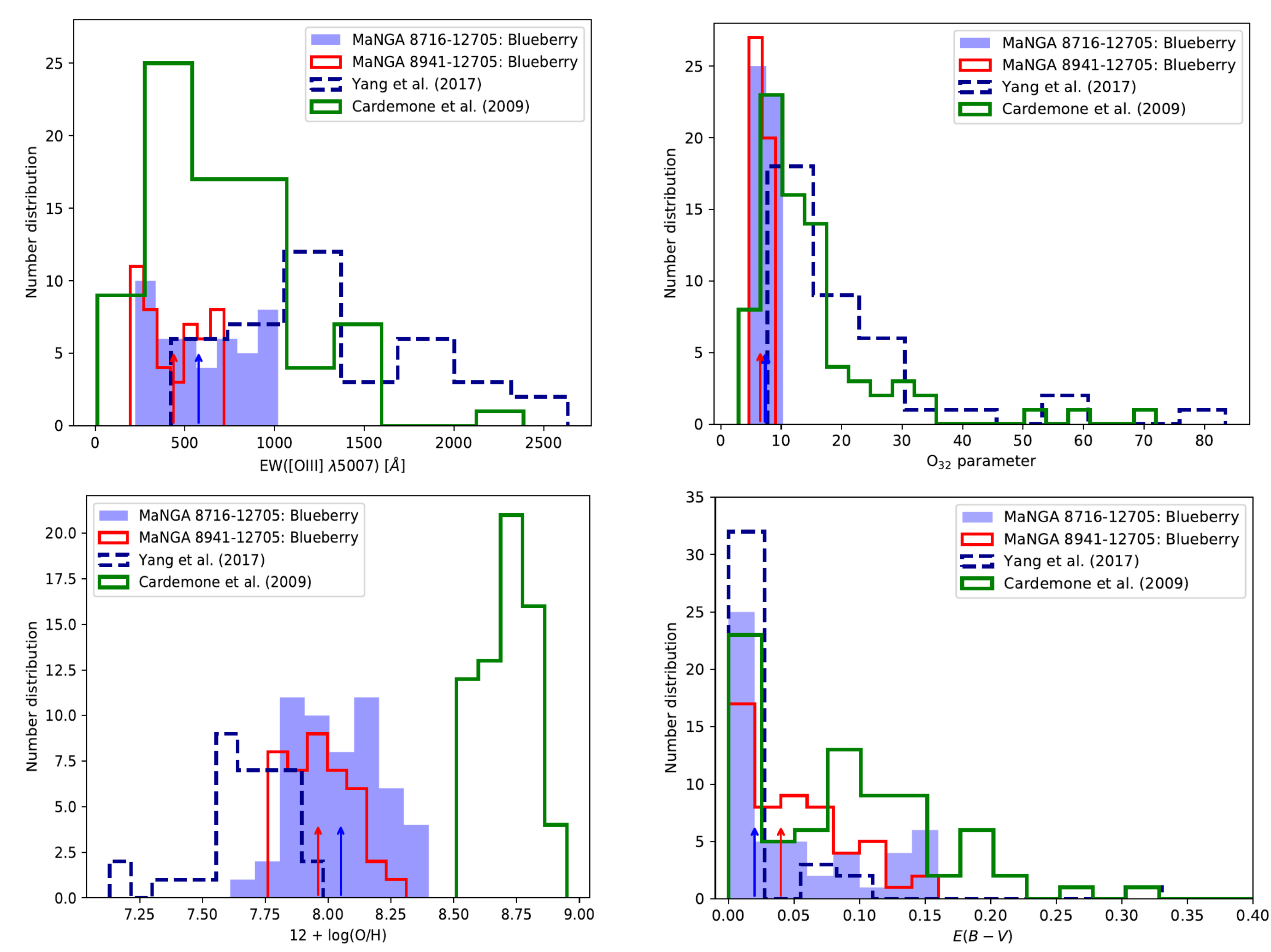}}
\caption{The spaxel histogram distribution of EW[OIII] (top-left), O$_{32}$ a proxy for ionization parameter (top-right), gas-phase metallicity [12 + log(O/H)] (bottom-left) and E($B - V$) (bottom-right) extracted over only blueberry region of each galaxy as marked by circles in Fig.~\ref{prop-map}. These distribution are also compared with the distribution of previously known blueberry and GP galaxies studied in \citet{Yang2017} and \citet{Cardamone2009}, respectively. In each plot, red (MaNGA: $8941 - 12705$) and blue (MaNGA: $8716 - 12705$) arrows show the median values of corresponding plotted parameters.}
\label{prop-dist}
\end{center}
\end{figure*}

The spatially resolved spectroscopic properties such as EW([OIII] $\lambda$5007), [OIII]/[OII] $\equiv$ O$_{32}$ parameter, dust extinction $E(B - V)$ and gas-phase metallicity 12 + log(O/H) of the selected compact H{\sc ii} regions as marked by red circle in the 2D IFU maps are presented in Fig.~\ref{prop-map}. In Fig.~\ref{prop-dist}, the same properties are also shown in the form of histogram distributions and then compared with the spectroscopic properties of previous known GPs and blueberry galaxies, respectively, studied in \citet{Cardamone2009} and \citet{Yang2017}. In this figure, GP and blueberry galaxies are shown by solid green and dashed blue histograms, respectively, while the spaxels from compact H{\sc ii} regions in the present work are represented by solid-red line (MaNGA: 8941-12705) and spacefilled-blue (MaNGA: 8716-12705) histograms. From Fig.~\ref{prop-map} and~\ref{prop-dist}, it is interesting to note that the derived spectroscopic properties of the selected compact H{\sc ii} regions are in consistent with properties of GP and blueberry galaxies - suggesting that these compact H{\sc ii} regions are found with extremely high values of EW([OIII] $\lambda$5007) and O$_{32}$ parameter, and with low values dust extinction and gas-phase metallicity, similar to typical spectroscopic properties of GPs and blueberries. Therefore, in consistent with our observed photometric color-color property that indicates the compact H{\sc ii} regions as blueberry candidates, our analysed spectroscopic properties also support the fact that the selected compact H{\sc ii} regions are indeed blueberry sources.\\   

\subsection{Sizes and projected distances}

Since the identified blueberry sources in photometric band images (see Fig.~\ref{color-comp}) suffer a possible faint and diffuse light contamination due their host LSB galaxies, it therefore makes difficult to define the actual extent or sizes of our blueberry sources. In order to follow a systematic approach and compare with the sizes of other extragalactic sources, we define the size of a given blueberry source as the area encircled the largest extent of H$\alpha$ emitting region seen in the 2D H$\alpha$ emission map centered at the peak emission associated with that region. This provides us an estimate of the total size of the blueberry region by defining an equivalent radius of $r_{equ}$ = $\sqrt{Area/\pi}$. Similarly, an effective radius ($r_{eff}$) is also defined as the radius that contains half of the total flux within the area covered by circle of radius $r_{equ}$. The values of $r_{equ}$ and $r_{eff}$ of our identified blueberry sources are listed in Table~\ref{Table1}, which range from a few hundreds of pc (i.e., $\sim$ 750 pc) to 2.1 kpc and $\sim$ 0.4 to 1 kpc, respectively. If we compare our estimated $r_{equ}$ with the sizes of other extragalactic H{\sc ii} regions studied in the literature \citep[e.g.,][]{Kennicutt1984,Mayya1994} which have typical radii of $\sim100 - 900$ pc, the present blueberry sources are comparable to the largest Gaint H{\sc ii} regions. Also our estimated $r_{eff}$ of the blueberry sources (i.e., $0.4 - 1$ kpc) is comparable to other typical Blue Compact Dwarfs (BCDs), Tidal Dwarf Galaxies (TDGs) and some of dwarf galaxies located in the Local Group as seen with $r_{eff}$ of 0.2 $\leq$ $r_{eff}$ $\leq$ 1.8 kpc \citep[e.g.,][]{Marlowe1997,Duc1998,Cairos2003,Mateo1998}. Moreover, sizes of the identified blueberries are consistent with other blueberries studied in the literature \citep[e.g.,][]{Yang2017}.\\

The projected distance between the host LSB galaxies and identified blueberries are provided in Table~\ref{Table1}, which range between $\sim$ 4 and 10 kpc. These estimated projected distances are relatively smaller, if we compare them with typical measured projected distances between TDGs and their parent galaxies available in the literature \citep[e.g.,][]{Duc1998}. Nevertheless, there are already several of identified TDGs \citep[e.g.,][]{Iglesias-Paramo2001,Amram2004} which are separated at smaller distances, similar to the present cases. Therefore, having the estimated sizes and projected distances, the identified blueberry sources situated at the outer most region of LSB galaxies most likely seem to be either tidally interacting systems such as TDGs or merging compact dwarf systems. This hypothesis is also supported by their stable dynamical structure as present in the next section. 

\subsection{Stability against internal motions}
\label{sec:stability}

\begin{figure}
\begin{flushleft}
\rotatebox{0}{\includegraphics[width=0.48\textwidth]{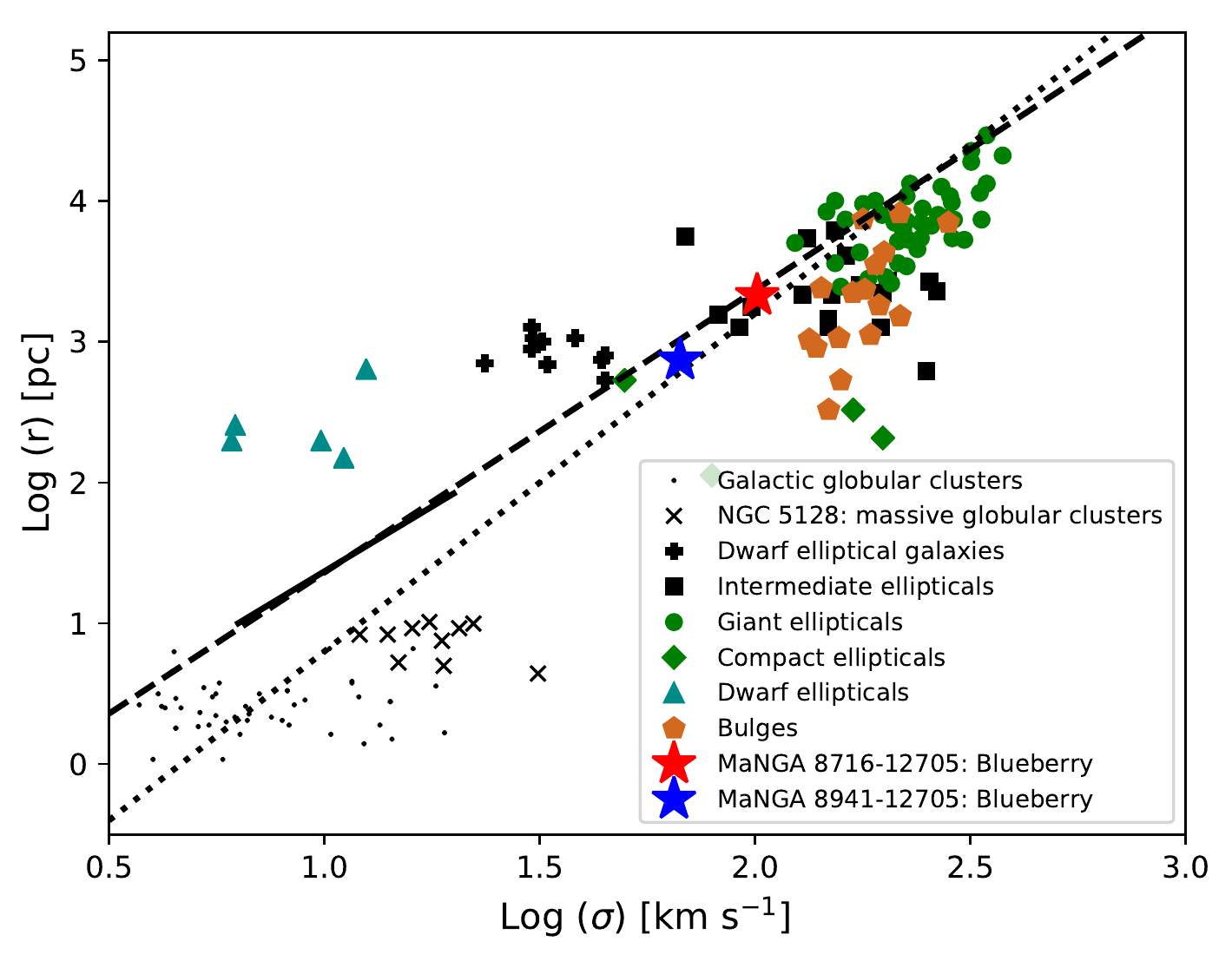}}
\caption{A comparison of our blueberries with different type of objects such as dwarfs, ellipticals, bulges on the $r_{equ} - \sigma$ plane (for details see the text). The blueberry galaxies in the present work are presented by star (red and blue) symbols.}
\label{r-sig}
\end{flushleft}
\end{figure}

In order to test that if the identified blueberry-like H{\sc ii} regions constitute themselves as self-gravitating entities to survive as a dwarf systems, we studied their locations on the radius-velocity dispersion (i.e., $r - \sigma$) relation measured for elliptical and globular clusters. For this purpose, we have plotted the classical fits to the $r_{equ} - \sigma$ plane for elliptical galaxies, globular cluster and H{\sc ii} regions taken from \citet{Terlevich1981} along with our identified blueberry sources as shown in Fig.~\ref{r-sig}. In this plot, different symbols represent the samples of various type of objects taken from the literature: dots $-$ galactic globular clusters from \citet{Trager1993}, \citet{Pryor1993}; crosses $-$ massive globular clusters in NGC 5128 from \citet{Martini2004}; plus $-$ dwarf elliptical galaxies from \citet{Geha2003}; square $-$ intermediate ellipticals; solid circles $-$ giant ellipticals; diamond $-$ compact ellipticals; triangles $-$ dwarf ellipticals and hexagonals $-$ bulges all these are from \citet{Bender1992}. Objects in this work are shown by star symbols. Here, the continuous line represents the fit for extragalactic H{\sc ii} regions taken from \citet{Terlevich1981} while the dashed and dotted lines show the fits for elliptical galaxies and globular clusters + elliptical galaxies, respectively. Based on the position of our blueberry sources on the $r_{equ} - \sigma$ plane, it appears that they are indeed self-gravitating entities. Furthermore, they are also in a region close to other dwarf and intermediate elliptical galaxies and much far from the regions occupied by globular clusters and extragalactic H{\sc ii} regions. Therefore, our analysis suggests that these blueberry sources in the present study are indeed typical compact dwarf systems which are associated with LSB galaxies either through a merger or tidal interaction event.   

\subsection{Main sequence relation for our blueberries}

\begin{figure}
\begin{flushleft}
\rotatebox{0}{\includegraphics[width=0.48\textwidth]{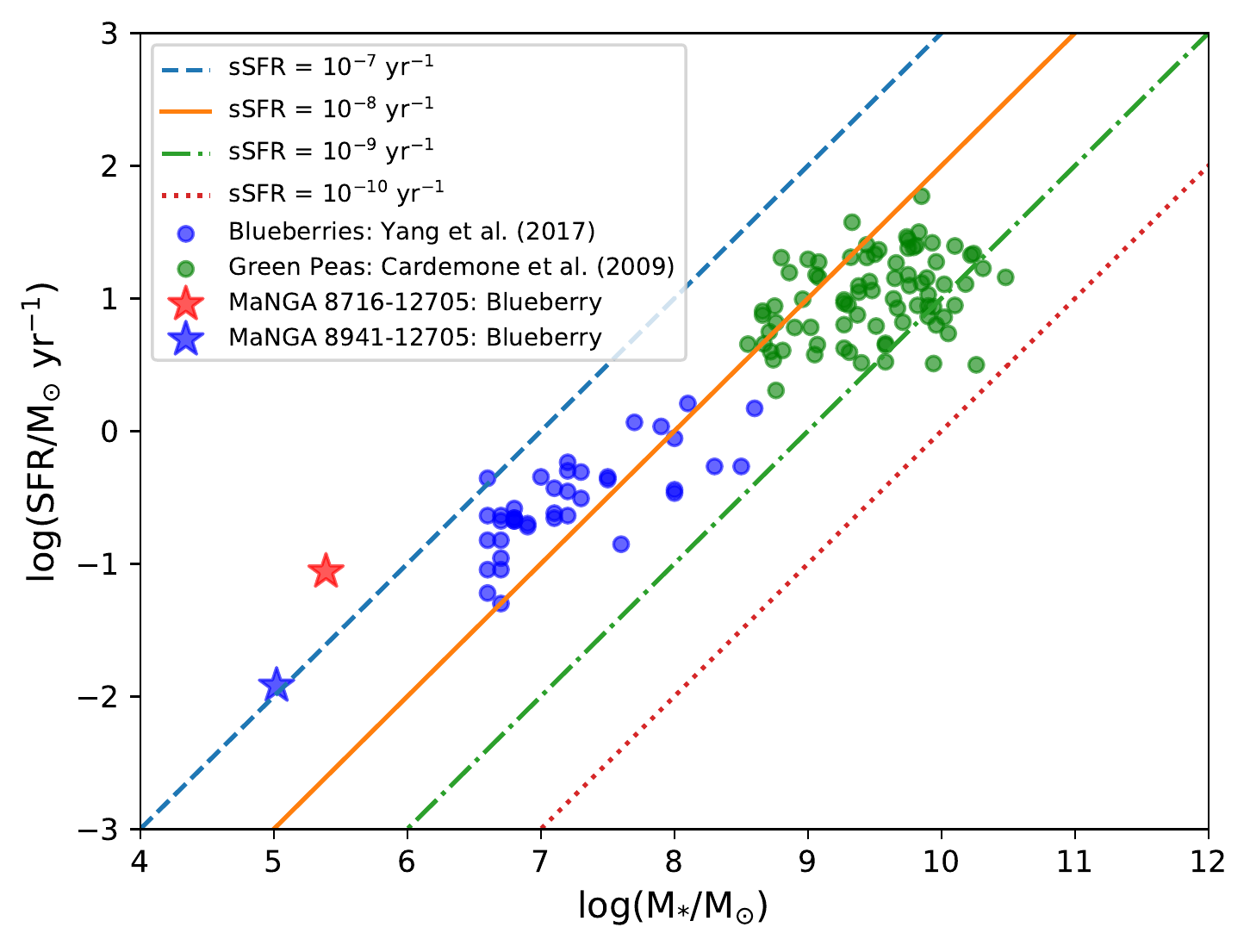}}
\caption{The stellar-mass vs. SFR for blueberry sources in the present study as represented by star symbol. This plot also includes other blueberry (blue circles) and GP (green circles) galaxies from \citet{Yang2017} and \citet{Cardamone2009}, respectively. Several straight lines in the plot represent constant sSFR of 10$^{-7}$, 10$^{-8}$, 10$^{-9}$ and 10$^{-10}$ yr$^{-1}$. }
\label{BB-MS}
\end{flushleft}
\end{figure}

In Fig.~\ref{BB-MS}, we have presented the main sequence ($M_{*}$ - SFR) relation for our identified blueberries (star symbols) together with other blueberries (blue circles) and GPs (green circles), respectively, taken from \citet{Yang2017} and \citet{Cardamone2009}. Here, SFRs are estimated from extinction corrected H$\alpha$ luminosity using SFR calibrator SFR(M$_{\odot}$~yr$^{-1}$) = 7.9 $\times$ 10$^{-42}$ $L_{H\alpha}$(erg~s$^{-1}$) provided by \citet{Kennicutt1998b}. Stellar-masses are estimated from ($g -r$) colors and r-band magnitudes using relation given by \citet{Bell2001}. The resulting SFRs and stellar-masses are provided in Table~\ref{Table1}. In Fig.~\ref{BB-MS}, dotted red line shows typical constant sSFR of 10$^{-10}$~yr$^{-1}$ observed for nearby normal star-forming galaxies in the $SDSS$ survey. It can be clearly seen that blueberries and GPs from \citet{Yang2017} and \citet{Cardamone2009}, respectively, are higher by $2 -3$ order of magnitude in sSFR than normal star-forming galaxies. This implies that the typical mass doubling time for GPs and blueberries are between 100 Myr and $\sim$ 1 Gyr. Interestingly, the identified blueberries in the present study are in consistent with these typical times. Furthermore, this also suggests that our observed blueberries are indeed starburst galaxies similar to other blueberries and GPs. Since stellar-masses and SFR of our blueberry galaxies are comparatively lower than that of other blueberries studied in \citet{Yang2017}, they therefore represent the extended smaller counterparts of GPs observed at relatively higher redshifts. 

\section{Gas-phase metallicity and its gradient}
\label{sec:metallicity}

\begin{figure*}
\begin{center}
\rotatebox{0}{\includegraphics[width=0.45\textwidth]{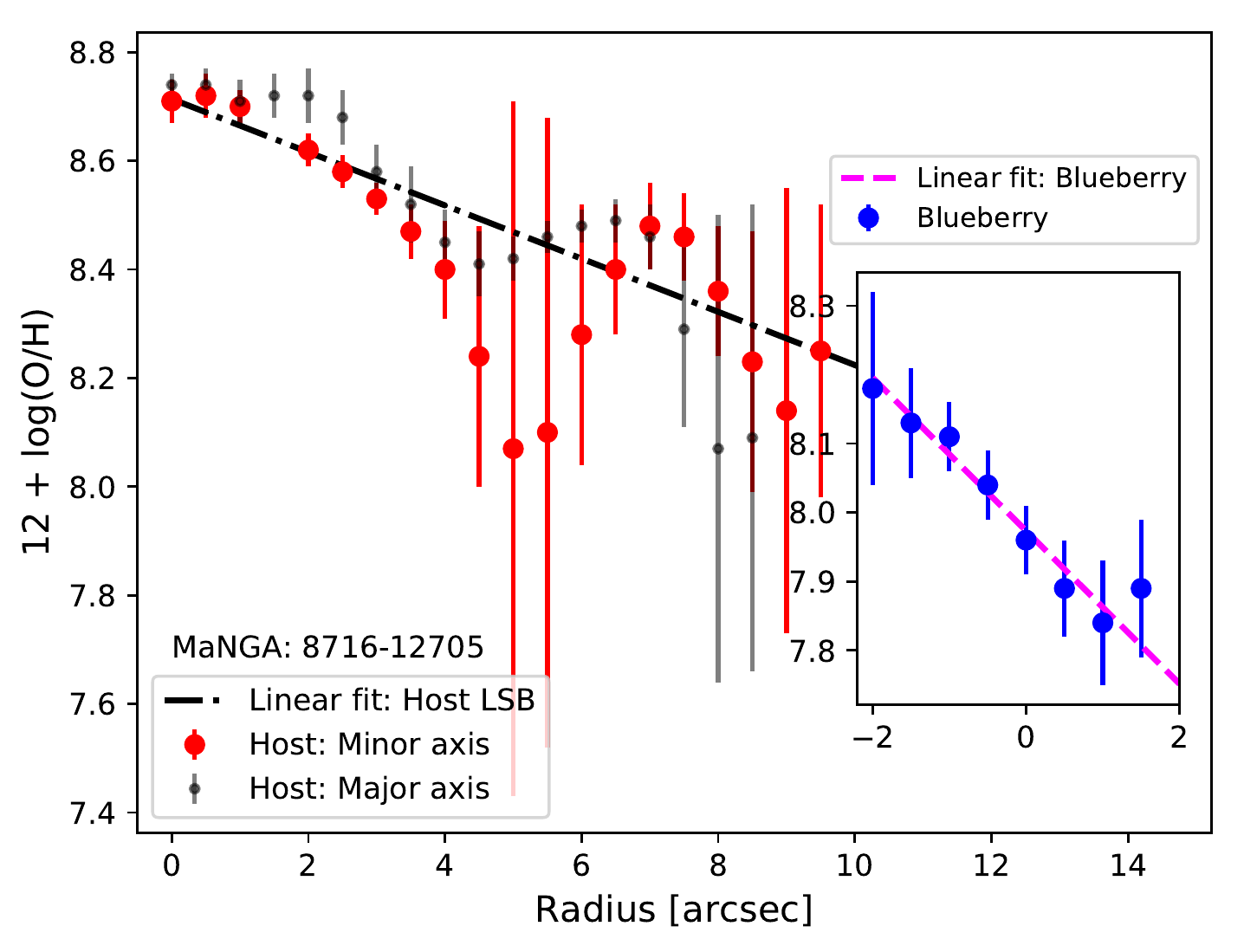}}
\rotatebox{0}{\includegraphics[width=0.45\textwidth]{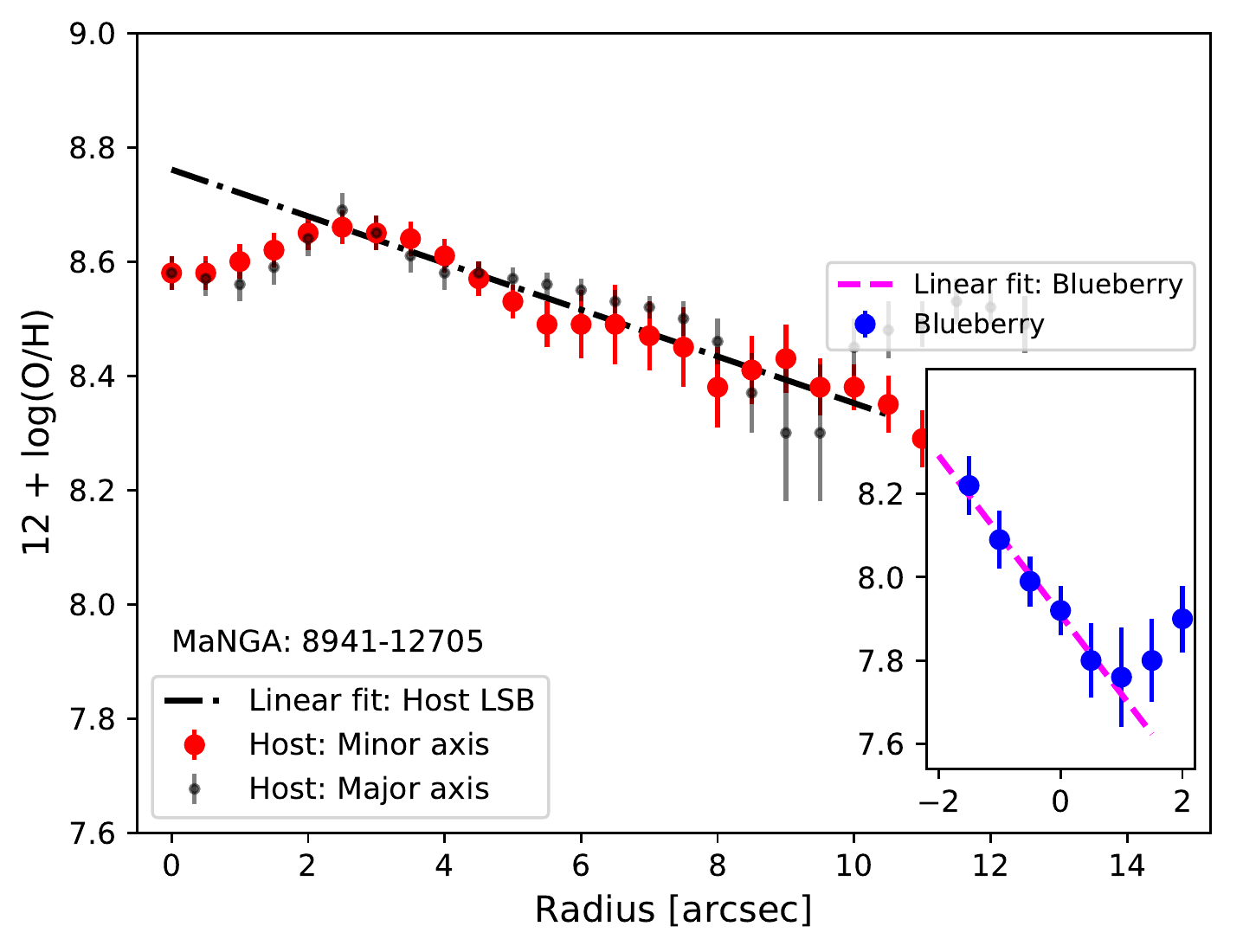}}
\caption{The radial gas-phase metallicity gradient in galaxies studied here. The red and blue solid circles correspond to the host LSB and blueberry galaxies, respectively. The dot-dsahed (black) and dashed (magenta) lines are linear-fit to the radial metallicity profiles in the host LSB and blueberry galaxies.}
\label{meta-prof}
\end{center}
\end{figure*}

In Fig.~\ref{meta-prof}, we have presented the radial profile of gas-phase metallicity for both the host LSB and blueberry galaxies. The gas-phase metallicity [i.e., 12 + log(O/H)] is derived using the N2 calibrator proposed by \citet{Pettini2004} - based on the ratio between the [NII]$\lambda$6583 and H$\alpha$ emission lines. In Fig.~\ref{meta-prof}, red and blue solid circles show the metallicity profiles for LSB and blueberry galaxies, respectively. Their respective linear fits to observed data are also shown by dot-dashed (black) and dashed (pink) lines. In this figure, black solid circles show the metalicity profile of LSB galaxies along their major axis (i.e., perpendicular to the profile shown by red solid circles which is drawn along the minor axis of LSBs, in the direction of identified blueberry galaxies).  \\

Interestingly, it is noticeable that the gas-phase metallicity gradients of LSB and blueberry galaxies are significantly different to each other. In each case, we found an appreciable mean metallicity difference of $\sim$ 0.5 dex between LSB and blueberry galaxies. Typical metallicity gradients in normal LSB spiral galaxies have been found between -0.015 and -0.033 dex~kpc$^{-1}$, with an average gradient of -0.024 dex~kpc$^{-1}$ having a scatter of 0.010 at 1$\sigma$ level \citep{Bresolin2015}. In consistent with the study of \citet{Bresolin2015}, our derived metallicity gradient for both the LSB galaxies under this study are found as $\sim$ -0.03 dex~kpc$^{-1}$. Nevertheless, the metallicity gradient of blueberry galaxies are found as -0.18$\pm$0.05 (MaNGA: 8941-12705) and -0.06$\pm$0.01 (MaNGA: 8716-12705), which are appreciably different than their respective host LSB galaxies, and also different than other normal LSB galaxies studied in the literature \citep[e.g.,][]{Bresolin2015}. Here, our observed mean metallicity difference of $\sim$ 0.5 dex and different metallicity gradient between LSB  and blueberry galaxies cannot be explained as a normal metallicity gradient as found in normal LSB galaxies studied in the literature. Alternatively, this inference suggests that the identified blueberry sources have experienced a different chemical enrichment history as compared to their respective host LSB galaxies, indicating that they are most likely different systems having significant different metallicities than their host LSBs. As evidence of metallicity difference within the galaxy has already been well-established as tool for identifying merging or tidally interacting systems \citep[e.g.,][]{Lopez-Sanchez2004a,Lopez-Sanchez2004b,Lopez-Sanchez2008,Lopez-Sanchez2010,Paswan2018}. Therefore, our blueberries studied here might be either TDGs or merging dwarf systems formed through tidal interactions or merger events, respectively.

\section{Possible Formation Scenario}
\label{sec:formation}

\subsection{Merger or TDG blueberry candidates?}

In previous sections, most of the observational properties of our studied blueberry galaxies such as sizes, projected distances to the host LSB galaxies and test for stability against their internal dynamics place them in the category of compact dwarf galaxies. Interestingly, the observed gas-phase metallicity gradient of blueberries are significantly different than their respective host LSB galaxies, and mean metallicity differences between blueberries and their host LSBs are also appreciably large (i.e., $\sim$0.5 dex). These observational evidences lead us to speculate that the identified blueberries may be either TDG candidates or merging dwarf systems. In this section, we have clarified our speculations as discussed below.\\

\begin{figure}[!h]
\begin{flushleft}
\rotatebox{0}{\includegraphics[width=0.48\textwidth]{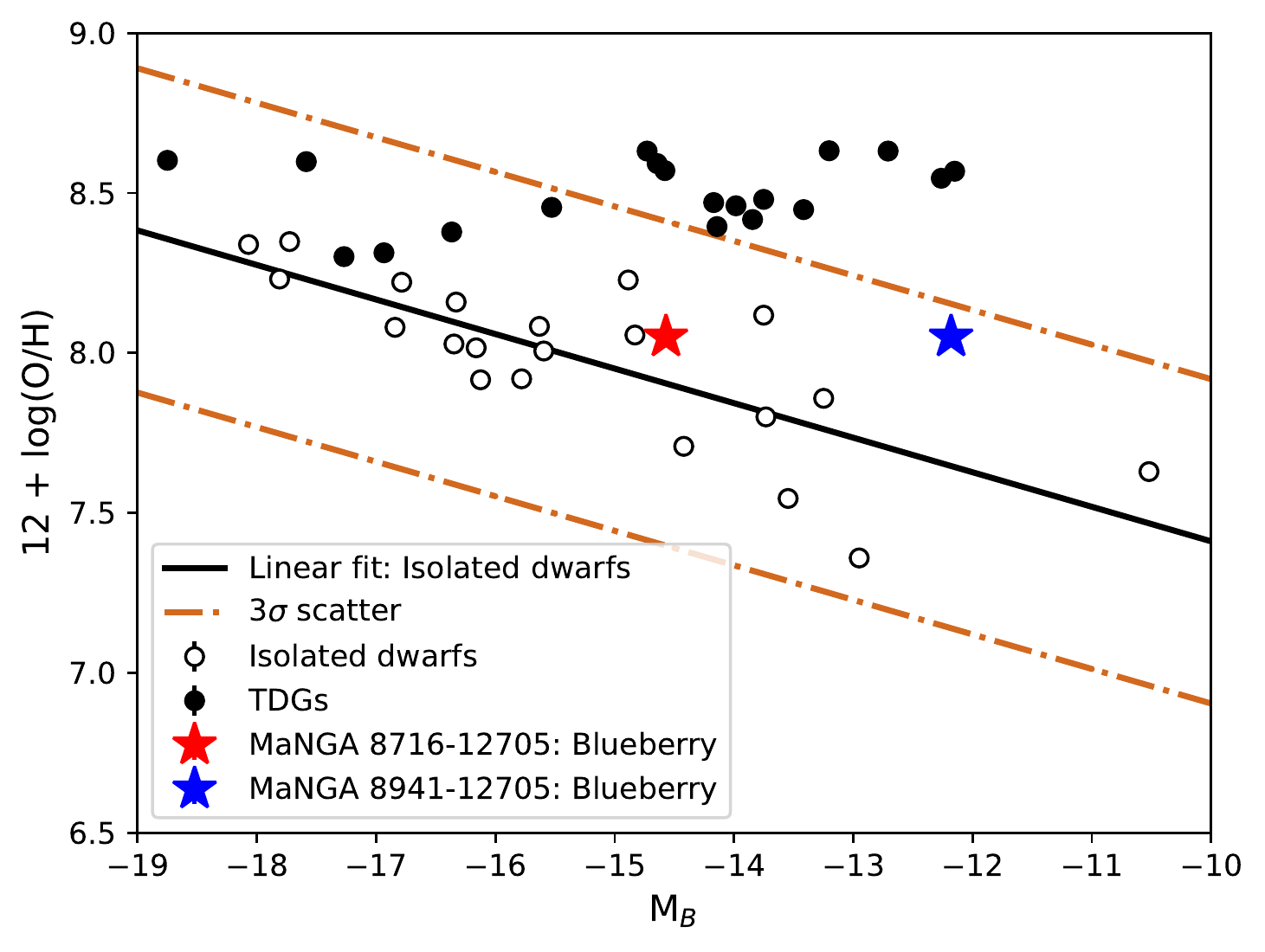}}
\caption{Oxygen abundance Vs. absolute $B$-band magnitude for a sample
of isolated nearby dwarf galaxies (open circles; taken from \citet{Richer1995} and
tidal dwarf galaxies (filled circles; taken from \citet{Duc1995}). Blueberry sources in the present work are shown with star symbols.}
\label{MB-Z}
\end{flushleft}
\end{figure} 

Our blueberry sources can be TDG systems, if they are enough massive to survive against the gravitational forces exerted by their nearest/host LSB galaxies. One approach to investigate whether our blueberries are stable against the gravitational potential of the host LSB galaxies is by using their tidal mass \citep{Binney1987, Mendes2001}, where the tidal mass is defined as: 

\begin{equation}
M_{tidal} = 3 M \left(\frac{R}{D}\right)^3
\label{Eq1}
\end{equation}

In Eq.~\ref{Eq1}, $M$ is the mass of host galaxy. $R$ and $D$ are the radius of blueberries (here used as $r_{equ}$) and their distance to the host LSB galaxies, respectively. If tidal masses estimated from Eq.~\ref{Eq1} are smaller than the masses of blueberry sources then only they are stable against the forces exercised by their host LSB galaxies, and hence they can survive as TDGs. The estimated tidal masses for each case are listed in Table~\ref{Table1}. In this table, it is to note that the estimated tidal masses in each case are found to be larger than the masses of blueberries, implying that they would not have any chance to be stable against the forces applied by LSB galaxies. Alternatively, this might also imply that they are in advanced stage of merger events.  \\

Moreover, other way to test the tidal origin of one dwarf system is the relation between absolute magnitude and gas-phase metallicity as proposed by \citet{Duc1998}. In Fig.~\ref{MB-Z}, M$_{B}$ vs 12 + log(O/H) relation is shown for normal dwarf galaxies (open circles) and TDGs (solid circles) taken from the literature. Here, blueberry sources in the current study are represented by star symbols. Since it is well-known that TDGs are made up from processed material in their parent galaxies \citep{Duc1998}, and hence they show high metallicity as also seen in Fig.~\ref{MB-Z}. In this figure, our studied blueberries are however in consistent with the locations of normal dwarf galaxies within 3$\sigma$ scatter (as shown by orange dot-dashed line) observed in the relation, showing low metallicity for a given absolute $B$-band magnitude of blueberry galaxies. Therefore, our analyses performed here rule out the fact that blueberries are TDGs, suggesting that they are indeed merging dwarf galaxies associated with LSB galaxies.   

\subsection{Local environment}

It is worth examining the local galaxy environment of our blueberries in a view of merger event as observed in this study. For this purpose, we searched for neighbor galaxies within a projected radius of 1 Mpc and having a narrow velocity range of nearly $\pm$ 250 km s$^{-1}$ from the recession velocity of host LSB galaxies. The reasons for selecting this parameter space are the following. A galaxy can travel up to $\sim$ 1 Mpc over a time of 10 Gyr (typical age of one galaxy) with a velocity of $\sim$ 250 km s$^{-1}$ in the IGM, which is also typical velocity dispersion in groups of galaxies. The local galaxy environment of each individual LSB/blueberry galaxies is briefly discussed below. 

\subsubsection{MaNGA: $8716 - 12705$}

Within the defined search parameters as mentioned above, a total of 39 galaxies are found in the vicinity of MaNGA: $8716 - 12705$. It gives an average galaxy density of $\sim$ 12 Mpc$^{-2}$. All the identified neighbouring galaxies are in a very narrow velocity range of $\sim 4100 - 4300$ km~s$^{-1}$, where the recession velocity of host LSB galaxy is 4207 km~s$^{-1}$. Our observed galaxy density value here is similar to typical group environments of galaxies as found in the literature \citep[e.g.,][]{Omar2005b,Sengupta2006}. This analysis therefore suggests the presence of a galaxy-rich dense environment around MaNGA: $8716 - 12705$, implying that this blueberry galaxy belongs to a dense group environment.  

\subsubsection{MaNGA: $8941 - 12705$}

In the vicinity of MaNGA: $8941 - 12705$, we found a total of 34 galaxies which gives an average galaxy density of $\sim$ 11 Mpc$^{-2}$. The neighbouring galaxies are in a narrow velocity range of $\sim 11980 - 12180$ km~s$^{-1}$, where the recession velocity of the host LSB galaxy is 12082 km~s$^{-1}$. Similar to previous case, it seems that MaNGA: $8941 - 12705$ also together with neighbouring galaxies forms a dense group of galaxies, and hence this blueberry source associated with LSB galaxy also belongs to a dense group environment. \\

With an estimated galaxy density of about $\sim$ 12 Mpc$^{-2}$ around blueberries studied here, indicating a likely higher interaction/merger rate, it is obvious that tidal interactions and/or merger events leading to intense starbursts are very common in this region.

\section{Summary and conclusions}
\label{sec:summary}

In this work, we studied two giant H{\sc ii} regions situated at the outer most region of LSB galaxies observed in the MaNGA survey. They are selected being the brightest H{\sc ii} regions in H$\alpha$ emission with the highest H$\alpha$ EWs among all the H{\sc ii} complexes seen in the LSB galaxies. They are also found to be closely associated with their host LSB galaxies. Further, the spatially resolved physical (e.g., optical color, size and distance to the LSBs, $r - \sigma$ relation, dust extinction, luminosity, EW, ionization parameter and main sequence relation etc.) and chemical (e.g., gas-phase metallicity and its gradient) properties of the selected H{\sc ii} regions are characterized using $SDSS$ photometric and MaNGA IFU spectroscopic data. Our main results are concluded as follows.

\begin{itemize}
    \item The two identified brightest giant H{\sc ii} regions at the outer most region of LSB galaxies are indeed compact dwarf galaxies ($1 -2$ kpc), as confirmed their locations in the region of typical dwarf galaxies on the standard $r - \sigma$ relation.
    
    \item Using photometric and spectroscopic characteristics, they are classified to be blueberry compact dwarf galaxies, similar to that studied in the literature.
    
    \item The studied blueberries follow the same main sequence relation as known for other GPs and blueberries in the literature. Nevertheless, they lie at lower mass end of this relation, indicating that they are extended fainter counterparts of other known blueberries and GPs. They also represent themselves as blueberry galaxies having the lowest stellar masses i.e., log($M_{*}/M_{\odot}$) $\sim$ 5 known to date.
    
    \item Our result supports the fact that the identified blueberry galaxies situated at the outer most region of LSB galaxies are in their advance stage of merger taking place in the dense galaxy environments. In contrast, other known blueberries in the literature are found to be resided in low density environments. Therefore, our blueberry galaxies here may represent a different population of blueberry-like systems which are formed through tidal interactions and/or mergers of low-mass (-luminous) galaxies or H{\sc i} clouds in a dense galaxy environment. 
\end{itemize}

This pilot study using data from the MaNGA survey demonstrates a great exploitation of IFU data to identify and characterize TDGs or merging compact systems located at outskirt of underlying galaxies in dense environments. Similar to this study, a systematic search using several other IFU surveys such as MUSE, CALIFA and SAMI can further enlarge our sample of different population of blueberry-like systems formed through interaction and/or merger of galaxies or H{\sc i} clouds. Moreover, the ultra-violet (UV) observations of such different sample of blueberries may identify them as Lyman Continuum (LyC) leaking sources responsible for re-ionizing our early Universe. Such study will help us in a broad understanding about the characteristics of LyC leakers resided in the diverse environments.  

\bibliography{ms-BB-TDGs}
\end{document}